\documentclass[twocolumn,showpacs,preprintnumbers,amsmath,amssymb,pre]{revtex4-1} 
\usepackage{color}

\usepackage{graphicx}
\usepackage{dcolumn}
\usepackage{bm}
\usepackage{subfigure}
\usepackage{amssymb}
\usepackage{tabularx}
\usepackage{multirow}

\renewcommand{\vec}{\mathbf}

\begin{document}

\title{Dynamics of two-dimensional complex plasmas in a magnetic field }

 \author{T.~Ott$^{1,2}$}
 \author{H.~L\"owen$^1$}%
\author{M.~Bonitz$^2$}%
 \affiliation{
 $^1$Institut f\"ur Theoretische Physik II: Weiche Materie, Heinrich-Heine-Universit\"at D\"usseldorf, Universit\"atsstra\ss{}e 1, D-40225 D\"usseldorf, Germany
\\$^2$Christian-Albrechts-Universit\"at zu Kiel, Institut f\"ur Theoretische Physik und Astrophysik, Leibnizstra\ss{}e 15, 24098 Kiel, Germany }
 
\date{\today}

\begin{abstract}
We consider a two-dimensional complex plasma layer containing charged dust particles
in a perpendicular magnetic field. Computer simulations of both one-component and binary systems are used to explore the
equilibrium particle dynamics in the fluid state.
The mobility is found to scale with the inverse of the magnetic field
strength (Bohm diffusion) for strong fields.
For bidisperse mixtures, the magnetic field dependence of the long-time
mobility depends on the particle species providing an external control of their
mobility ratio.
For large magnetic fields, even a two-dimensional model porous matrix
can be realized composed by the almost immobilized high-charge particles
which act as obstacles for the mobile low-charge particles.

\end{abstract}

\maketitle

\section{Introduction}

Transport properties in liquids are relevant for various applications
ranging from solvation of tablets~\cite{Colombo1999}
the penetration of salt ions into fresh river water~\cite{Wright1971} to
imbibition problems~\cite{Gruener2012}. Hence,
there is a need for a basic understanding of particle diffusion on the most fundamental level of the individual
particles. The particle trajectories, as governed by Newton's equation of
motion with the interparticle forces,
are the natural starting point to understand and predict the transport
properties~\cite{Voigtmann2009,*Perera1999,*Winkler2004}.
Already in equilibrium this is still a nontrivial problem of classical
statistical mechanics.

Complex plasmas~\cite{Bonitz2010,Morfill2009,Ivlev2012} which contain
mesoscopic dust grains are ideal model systems to follow the dynamics on the time and length scale of the
individual particles.
Typically the dust particles are highly charged such that there are strong
repulsive effective interactions between neighbors. At high densities, the system can therefore exhibit both fluid
and solid phases~\cite{Thomas1994}. In this paper, we study the particle dynamics in a two-dimensional complex
plasma which is exposed to an external magnetic field of strength $B$.
Though the presence of a magnetic field does not alter the equilibrium static
properties, such as phase transitions, the dynamics are strongly affected~\cite{Ott2013}. 
Due to the Lorentz force, the charged particles exhibit a circular motion~\cite{Kahlert2012,Ott2011c,Bonitz2013}
which is expected to slow the dynamics down. Therefore the magnetic field opens the fascinating
possibility to change the dynamical properties of the system externally without changing the system itself, e.g., Ref.~\cite{Hartmann2013}.

\begin{figure}[ht]
\centering \includegraphics{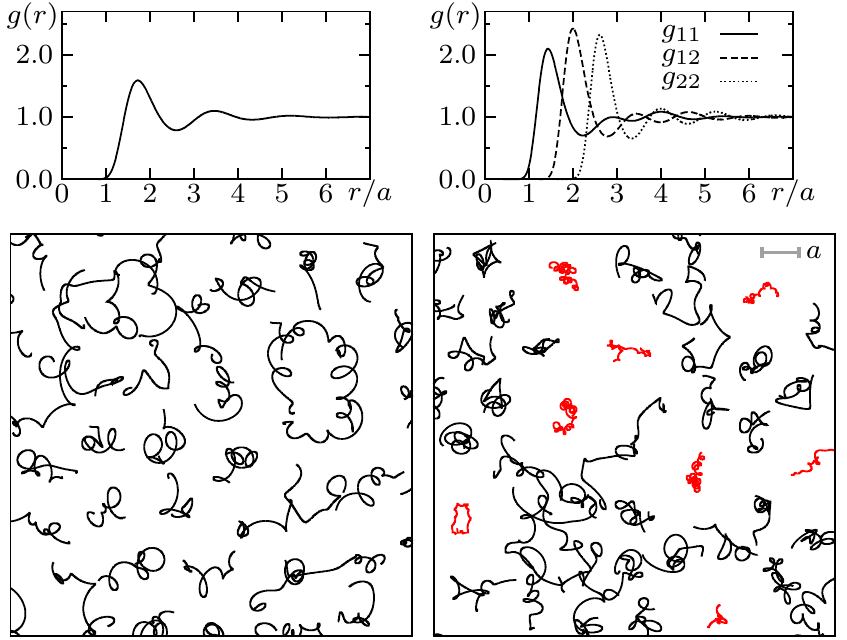}
\caption{(Color online) Pair distribution functions and particle trajectories during $\omega_p t=30$ at $\Gamma=30$, $\beta=0.5$. 
\emph{Left:} One-component system. \emph{Right:} Binary mixture with $Q_r=4$ and $n_r=1/3$. The highly charged 
particles are shown in red/light. \label{fig:traj_ex}}
\end{figure}

Some aspects of the one-component dust particle dynamics in a magnetic field have
been considered in previous studies~\cite{Ott2011c,Ott2012,Hou2009c,Ott2013,Ott2013a,Bonitz2010,Ott2011,*Ott2011a}.
In particular, the motion of few-particle clusters in magnetic fields
has been studied both by experiments and simulation~\cite{Kahlert2012,Bonitz2013}. 
The dynamics of three-dimensional ionic binary mixture have also been under investigation in early 
as well as recent research~\cite{Hansen1985,Bastea2005,Daligault2012,Apolinario2011}, with a particular focus on astrophysical 
plasmas and those encountered in inertial confinement fusion.

 Here, we focus on two-dimensional systems and explore the long-time
dynamics by computer simulations.
 We confirm the $1/B$-scaling of the long-time diffusion coefficient
 for strong magnetic fields~\cite{Ott2011c} for the two-dimensional system. 
 We moreover consider binary systems composed of
 high-charge and low-charge particles~\cite{Assoud2008,Dzubiella2002,Gray2001}. Our motivation to study a
 binary system comes from the fact that the magnetic field affects the
dynamics of the particle species differently. Thereby, the individual particle
dynamics can be steered externally via the magnetic field. One important parameter is the
mobility ratio of the two species which governs the mutual diffusion and is one key parameter
 for the nature of the kinetic glass transition in mixtures~\cite{Imhof1995,Zaccarelli2004,Mayer2008}.
 This ratio is typically fixed by the mass ratio~\cite{Barrat1990} and the
interactions~\cite{Assoud2009} and can therefore not easily be tuned. Here we show that this ratio can
be controlled by an external magnetic field insofar as the high-charge particles are more
immobilized than the low-charge particles. For large magnetic fields, it is even conceivable
that the high-charge particles are almost immobilized while the low-charge particles are still
mobile. This opens the way to realize a porous model matrix in two dimensions. Recently a
 similar matrix has been created by adsorbing colloidal particles to a
substrate~\cite{Deutschlander2013}.
 Our approach, however, is more flexible as everything can be controlled
externally.

This paper is organized as follows: in section~\ref{sec:model} we describe our model of complex plasmas in  a
magnetic field. In \ref{sec:homsys} we describe results for the one-component case. The binary
mixtures is then considered in section~\ref{sec:binsys}. Finally, we conclude in section~\ref{sec:summary}.

\begin{figure}[t]
\centering \includegraphics{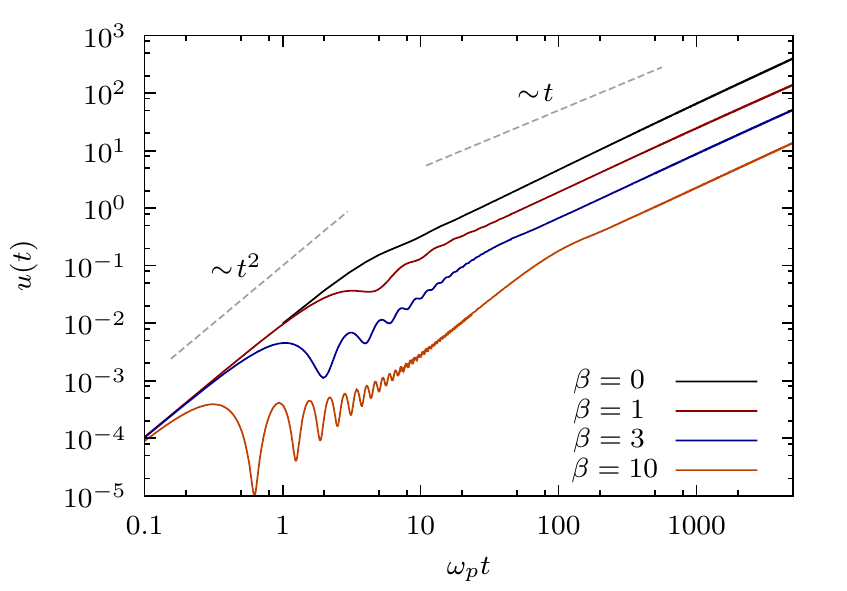}
\caption{(Color online) MSD of a one-component system with $\Gamma=100$ at different magnetic 
field strengths.The straight lines indicate linear and quadratic growth.  }
\label{fig:msd_k1g100qr1}
\end{figure}

\begin{figure}[t]
\centering \includegraphics{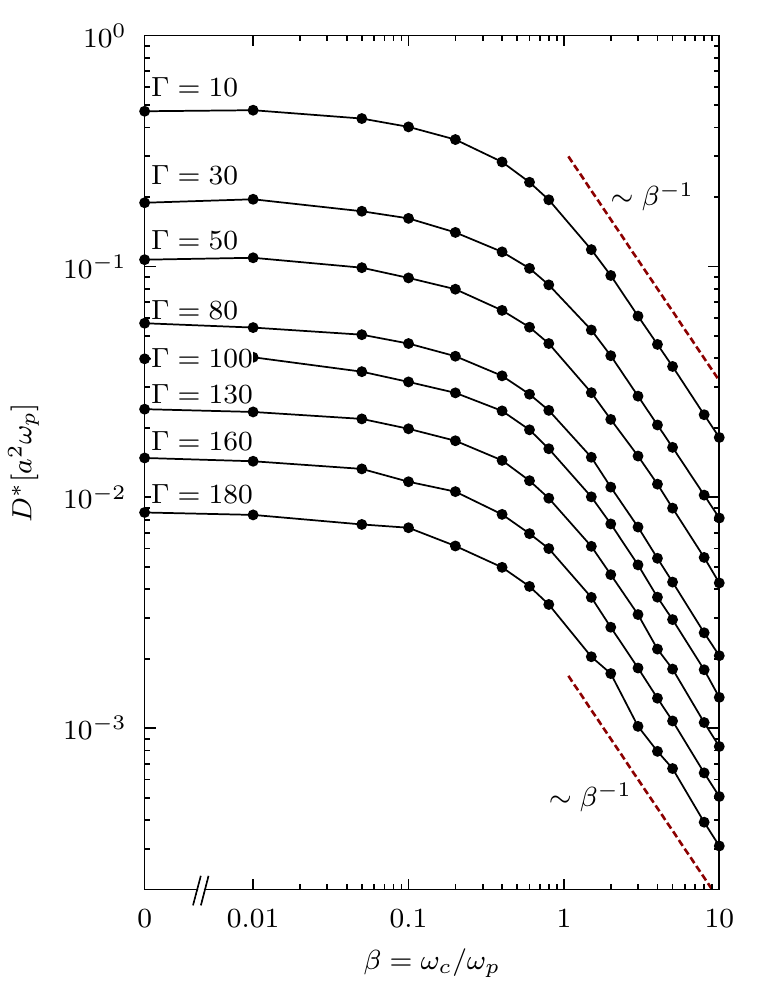}
\centering \includegraphics{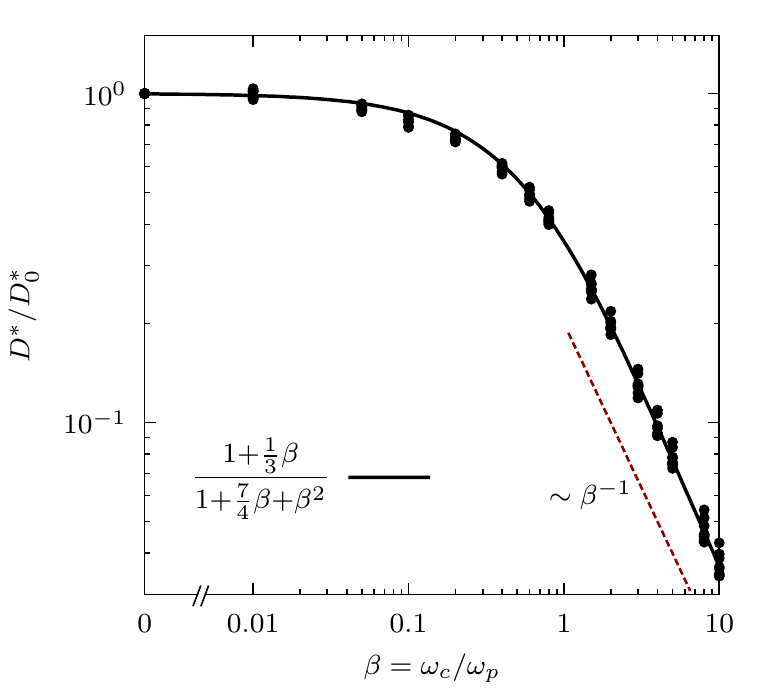}
\caption{(Color online) \emph{Top:} $D^\ast$ as a function of $\beta$ for values of $\Gamma$ as indicated in the figure. The 
dotted lines show a decay $~\beta^{-1}$ as a guide for the eye. \emph{Bottom:} $D^\ast$ normalized by the field free value $D^\ast_0=D^\ast(0)$. The 
normalized values fall on a universal curve for all values of $\Gamma$. }
\label{fig:dcoeff_b}
\end{figure}

\section{Model}
\label{sec:model}

We consider one-component systems and charge-asymmetric binary mixtures of uniform mass $m$, charge ratio $Q_r=q_2/q_1$, and 
density ratio $n_r=n_2/n_1$, where the numeric indices label the particles species. The particles are situated in a two-dimensional quadratic simulation 
box of side length $L$, giving rise to partial densities $n_{1,2}=N_{1,2}/L^2$, and interact via a screened Coulomb interaction with screening length $\lambda$, 
\begin{align}
 V_{ij}(\vec r_i, \vec r_j) &= \frac{q_iq_j}{\vert\vec r_i - \vec r_j\vert}\exp\Big(-{\vert\vec r_i - \vec r_j\vert/\lambda}\Big).
\end{align}

In addition, we consider the influence of an external magnetic field $B$ perpendicular to the particle plane, giving rise to the cyclotron 
frequency $\omega_{c,1,2}=\vert q_{1,2}\vert B/(mc)$ ($c$ is the speed of light). 

The system is fully described by a set of five parameters: $Q_r$, $n_r$, $\kappa$, $\Gamma$, and $\beta$. Here, the screening strength is 
defined as $\kappa=a/\lambda$ with the Wigner-Seitz radius $a=(\pi(n_1+n_2))^{-1/2}$, $\Gamma=\Gamma_1=Q_1^2/(ak_BT)$ ($T$ is the temperature), and $\beta=\beta_1=\beta_2=\omega_{c,1,2}/\omega_{p,1,2}$, 
where $\omega_{p,1,2}=\big(2q_{1,2}^2/(a^3m)\big)^{1/2}$ is the nominal Coulomb plasma frequency. In the following, we normalize lengths by $a$ and 
times by the inverse of $\omega_p\dot = \omega_{p,1}$.

Our investigations are carried out by molecular dynamics simulation for $N=16\,320$ particles and encompass a measurement time of $\omega_pt=100\,000$ which 
is preceded by an equilibration period. The simulation is carried out in the microcanonical ensemble at $\kappa=1$; typical trajectory snapshots are shown 
in Fig.~\ref{fig:traj_ex}. Notice the familiar circular paths induced by the magnetic field and the different mobility of the particle species 
in the binary system. An external magnetic field does not influence the equilibrium structure of one-component systems or binary 
mixtures (Bohr-van Leeuwen theorem). The charge ratio, on the other hand, has a strong influence on the structure as quantified by the pair distribution function $g_{\alpha\beta}(r)$, see upper graphs in Fig.~\ref{fig:traj_ex}. In the binary mixture, the lightly charged particles exhibit a smaller correlation gap 
at small distances and a lower peak height, indicating a smaller degree of correlation in this subsystem. The highly charged subsystem is considerably more 
correlated (see $g_{22}(r)$ in Fig.~\ref{fig:traj_ex}), and the cross-correlation between the particles species ($g_{12}(r)$) falls in-between. 

The study of the dynamics of the system is undertaken by calculating the mean-squared displacement (MSD) $u(t)$ defined as
\begin{align}
u(t) &= \langle \vert \vec r_i(t) - \vec r_i(t_0) \vert^2 \rangle_{i, t_0}, 
\end{align}
where the averaging is over all particles and all starting times $t_0$. According to classical transport theory, the diffusion coefficient 
follows as
\begin{eqnarray}
 D &= \frac{1}{4} \lim\limits_{t\rightarrow \infty}\frac{u(t)}{t}.\label{eq:dcoeff}
\end{eqnarray}
Since the existence of Fickian diffusion is doubtful for strongly coupled two-dimensional Yukawa systems~\cite{Donk'o2009, Ott2009b, *Ott2009c}, 
we evaluate Eq.~\eqref{eq:dcoeff} at a fixed time instant $t\omega_p=4850$ and denote it $D^\ast$, keeping in mind that this measure of the mobility should not be identified with the long-time diffusion coefficient.

\section{One-component system}
\label{sec:homsys}
Before investigating the binary system, we first establish the general diffusion trends in a magnetized, one-component 2D Yukawa system, which are laid out here for the first time~\footnote{The magnetized 2D Coulomb ($\kappa=0$) system was first investigated in Ref.~\cite{Ranganathan2002}. Similar results have been obtained by Z. Donk\'{o}~\cite{Donko2009}}.~\nocite{Ranganathan2002,Donko2009}
The behavior of the MSD in such a system at $\Gamma=100$ is shown in Fig.~\ref{fig:msd_k1g100qr1} for different 
magnetic field strengths. For $\beta=0$, the ballistic regime with a quadratic increase at small times is followed by a quasi-diffusive regime 
in which the MSD grows almost linearly with time. With increasing magnetic field, the signature of the circular paths is visible in the MSD curves 
as an oscillatory growth. 
The localization of individual particles 
at high magnetic field values gives rise to an additional regime at very small time delays during which the MSD is subdiffusive, i.e, during 
which $u(t)$ grows less than linearly with time. 

The scaling of the diffusivity $D^\ast$ as a function of the magnetic field strength is of central interest with regard to the dynamics of the system. This scaling is shown in Fig.~\ref{fig:dcoeff_b}. For small values of $\beta$, the rapidity of the diffusive motion is unaffected, regardless of the coupling constant $\Gamma$. Only when magnetic field effects become important at $\beta \gtrsim 0.1$ does $D^\ast$ begin to decay. At $\beta \approx 1$, the scaling becomes the familiar Bohm type diffusion, $D^\ast\propto 1/\beta$~\cite{Spitzer1960}. This is the same behavior that was found in the diffusion \emph{perpendicular} to the field in strongly coupled three-dimensional OCPs~\cite{Ott2011c}.

\begin{figure}[t]
\centering \includegraphics{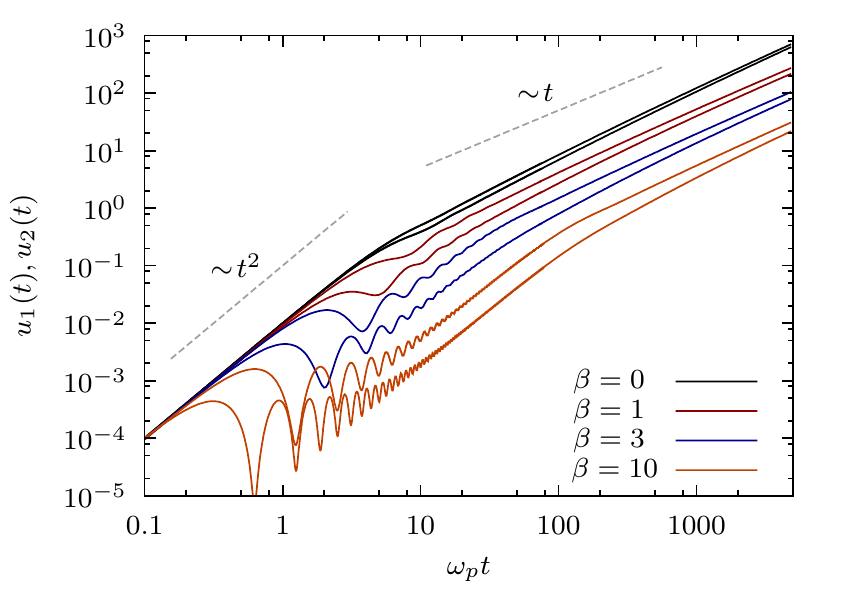}
\caption{(Color online) MSD of a binary system with $n_r=1$, $Q_r=0.5$, and $\Gamma=100$ at different magnetic 
field strengths. The lower of each pair of curves corresponds to the more highly charged particles. The straight lines indicate linear and quadratic growth. }
\label{fig:msd_k1g100_qr05}
\end{figure}

The functional form of the $D^\ast(\beta)$ dependence is quite insensitive to $\Gamma$, as demonstrated in the lower graph of Fig.~\ref{fig:dcoeff_b}. This is in contrast with the corresponding behavior of a three-dimensional OCP~\cite{Ott2011c} which shows a clear $\Gamma$-dependence both in field-parallel and perpendicular diffusion. The reason for the more complex behavior in 3D systems is the mutual interference between the two diffusion directions (mediated by the strong coupling between the particles), which is absent in 2D systems. 

\section{Binary system}
\label{sec:binsys}

\begin{figure}[t]
\centering \includegraphics{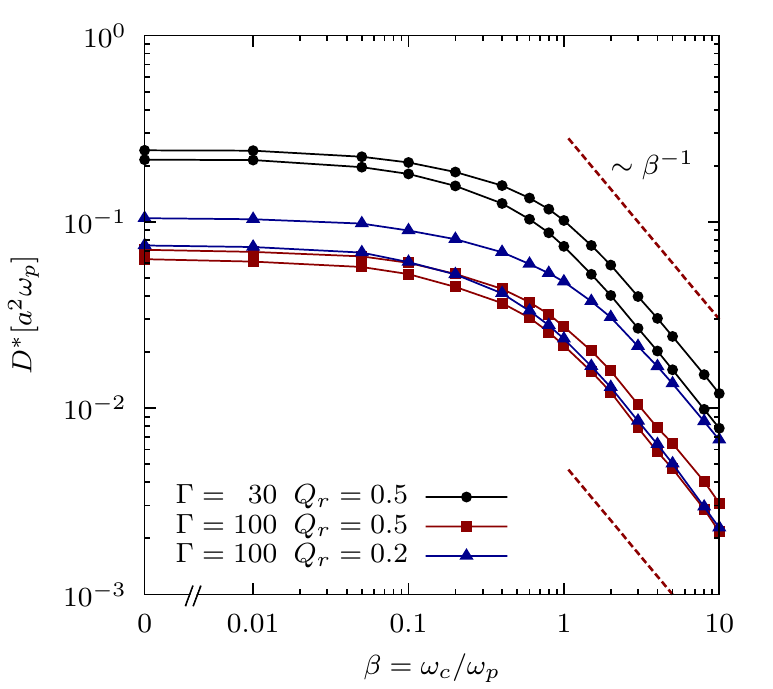}
\caption{(Color online) $D^\ast$ as a function of $\beta$ for binary systems with charge ratio $Q_r=0.5$ and $Q_r=0.2$. The lower one of each pair of curves corresponds to the more highly charged species.  The 
dotted lines show a decay $~\beta^{-1}$ as a guide for the eye. }
\label{fig:dcoeff_b_qr}
\end{figure}

In this section, we expand on the previous investigation and consider charge-asymmetric binary Yukawa systems with 
a repulsive interaction. The density ratio is fixed to $n_r=1$, i.e., $N_1=N_2$, while the charge ratio $Q_r$ and the magnetic field strength $\beta$ are varied. 

Figure~\ref{fig:msd_k1g100_qr05} shows the MSD of such a binary system at $Q_r=0.5$ and different magnetization; for each value of $\beta$, there are two curves, reflecting the two particle species. Evidently, the particles carrying a lower charge are more mobile, regardless of the magnetic field strength. For increasing $\beta$, however, the disparity in mobility between the two species grows steadily, as evidenced by the increasing gap between the two MSD curves when going from zero magnetic field to $\beta=10$. 

More data are presented in Fig.~\ref{fig:dcoeff_b_qr} which shows $D^\ast$ as a function of $\beta$ for two values of $Q_r$. The functional form of the data is comparable to the one-component case considered in the previous section and is well described by Bohmian diffusion for both species for $\beta\gtrsim 1$. A closer look, however, reveals 
that the response of the less highly charged (more mobile) species is shifted to higher values $\beta$ which results in an increase in the mobility ratio between the two species.  

\begin{figure}[ht]
\centering \includegraphics{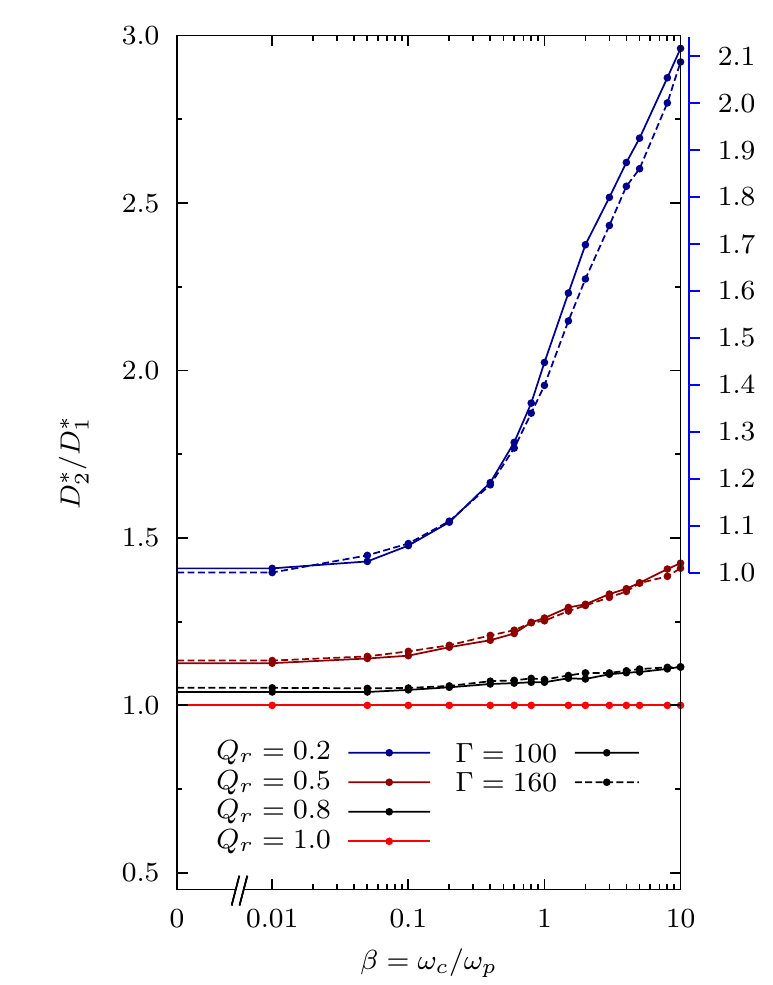}
\centering \includegraphics{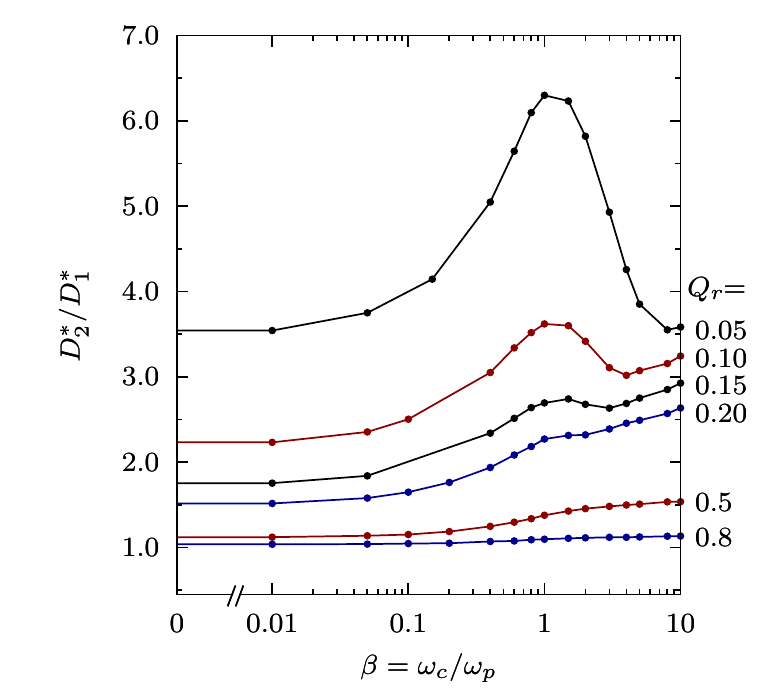}
\caption{(Color online) Ratio  $D_2^\ast/D_1^\ast$ for $\Gamma=100;160$ (top) and $\Gamma=30$ (bottom) at different charge ratios $Q_r$
as a function of $\beta$. The right axis in the top graph shows the relative change for $Q_r=0.2, \Gamma=100$, normalized to $\beta=0$. }
\label{fig:dratio_binary}
\end{figure}

In the upper graph of Fig.~\ref{fig:dratio_binary}, this is demonstrated for strong coupling, $\Gamma=100;160$, by plotting the ratio $D^\ast_2/D^\ast_1$. While a modest charge ratio of $Q_r=0.8$ results only in a small variation of this ratio, the influence of the magnetic field grows with decreasing $Q_r$, so that at $Q_r=0.2$, the magnetic field alone can be used to manipulate 
the mobility ratio by a factor of two for $\beta=10$ (see right-hand scale in Fig.~\ref{fig:dratio_binary}). Since the mobility ratio plays a crucial role during the glass transition, this effect can be leveraged to investigate the conditions for glass formation in one and the same system by controlling the mobility ratio by the external magnetic field. 

The relatively simple, monotonic dependence of $D^\ast_2/D^\ast_1$ on $\beta$ for strongly coupled plasmas shown in the upper part of Fig.~\ref{fig:dratio_binary} has to be contrasted with the more intricate behavior of the same ratio for $\Gamma=30$ (lower graph in Fig.~\ref{fig:dratio_binary}). Here, a highly non-monotonic 
dependence of the ratio $D^\ast_2/D^\ast_1$  is observed, which becomes more strongly pronounced for more disparate charge ratios.
 The strong growth of 
$D^\ast_2/D^\ast_1$, which unfolds undisturbed at large values of $\Gamma$, is suppressed at magnetic field strengths surpassing $\beta_c\approx 1$, leading 
to the formation of a pronounced peak at $\beta_c$. 

\begin{figure}[ht]
\centering \includegraphics{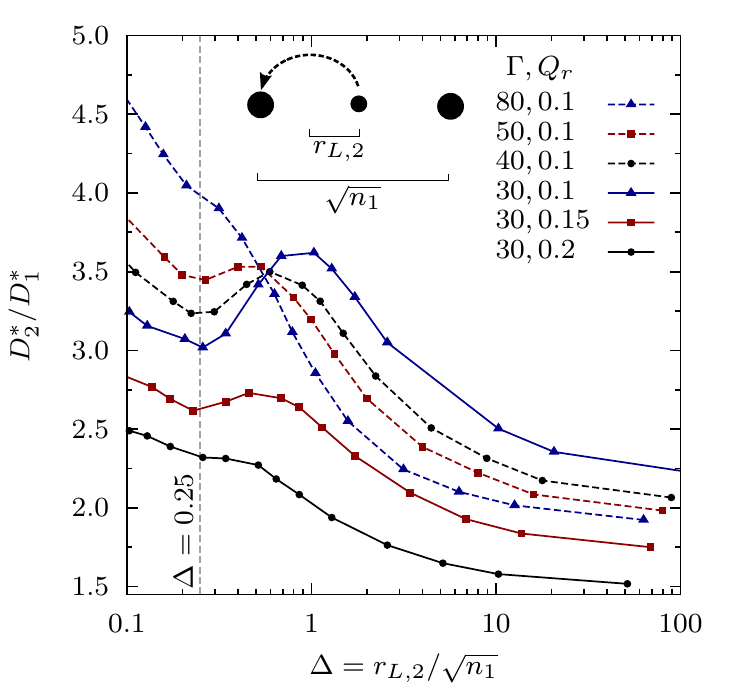}
\caption{(Color online) Ratio  $D_2^\ast/D_1^\ast$ as a function of $\Delta=r_{L,2}/\sqrt{n_1}$. Note that small values of $\Delta$ 
correspond to large magnetic fields, and vice versa. }
\label{fig:dratio_delta}
\end{figure}

The microscopic reason for the suppression at large magnetic fields lies in the 
reduced mobility of the lightly charged species. In a qualitative way, it can be traced back to a geometric origin 
by considering the ratio of length scales $\Delta=r_{L,2}/\sqrt{n_1}$ between the Larmor 
radius $r_{L,2}$ of the lightly charged species and the average nearest-neighbor distance 
$\sqrt{n_1}$ of the highly charged species. A lightly charged particle situated between two highly charged ones will be forced on a curved 
trajectory by the magnetic field. At $\Delta=1/4$, this trajectory leads, at thermal velocity of the particle, to a collision with one of the highly charged
particle, effectively preventing the diffusion of the lightly charged particle (see schematic in Fig.~\ref{fig:dratio_delta}). This 
results in a reduction of $D^\ast_2/D^\ast_1$ at $\Delta=1/4$. We have tested this simple geometric reason
by plotting data for the mobility ratio as a function of the geometric parameter $\Delta$, see Fig.~\ref{fig:dratio_delta} (notice that $r_{L,2}$ is a function of both $\Gamma$ and $Q_r$). 
In fact, a resonant dip in the mobility ratio is formed around $\Delta=1/4$
supporting the underlying picture.
This geometric resonance effect persists across different parameter regimes, 
but becomes less pronounced as $Q_r$ or $\Gamma$ are increased, 
since an increased particle coupling leads to stronger caging effects.

 \section{Conclusion}
 \label{sec:summary}

In conclusion, we have explored the dynamics of charged particles in a
complex plasma layer exposed to a perpendicular magnetic field which allows for an additional
external control parameter for the particle transport.
Our simulation results can be verified in experiments of dusty plasmas
either in magnetic fields~\cite{Thomas2012} or in rotating electrodes which formally lead to the same
equations of motion~\cite{Kahlert2012,Bonitz2013,Hartmann2013}.
Binary systems can also be realized in dusty plasmas, e.g., Ref.~\cite{Hartmann2009}. 

We have demonstrated that the mobility in two-dimensional systems adheres to the same $1/B$-Bohm scaling as 
in three-dimensional systems. In contrast to three-dimensional systems, however, the functional form of the scaling 
is largely independent of the coupling $\Gamma$, indicating a decoupling of magnetic and interaction effects. 

Our main focus has been on the response of a charge-asymmetric binary mixture to an external magnetic field. Since the two 
subsystems are affected differently by the magnetic field, the mobility ratio between them can be controlled by the strength of the magnetic field. 
For less strongly coupled systems and high charge-asymmetry, we have found that the circular trajectories of the lightly charged particles 
can be in resonance with the positional configuration of the highly charged particles, which leads to a distinct reduction of the mobility of 
the former. This is an interesting realization of a porous model matrix in a fluid system.

For future studies, as regards binary systems, a systematic understanding of the two-dimensional glass transition in binary mixtures is lying ahead where
the magnetic field is exploited as a steering wheel to change the mobility ratio between the particle species. 
Moreover, it is known that the crystallization process out of an undercooled melt 
depends sensitively on the mobility ratio  in binary systems
\cite{Loewen2012} such that the magnetic field can be used to tune 
crystal nucleation in mixtures \cite{Assoud2011,Loewen2011,Taylor2012,Ganagalla2013}.

\begin{acknowledgments}
We thank Zoltan Donk\'o and Peter Hartmann (Budapest) for numerous stimulating discussions in the early stages of this work. 
This work is supported by the Deutsche Forschungsgemeinschaft via SFB TR 6 and SFB TR 24 and grant shp0006 
at the North-German Supercomputing Alliance HLRN. 
\end{acknowledgments}


%

\end{document}